\documentclass[aip,jcp,showpacs,showkeys,preprint,eqsecnum,tightenlines]{revtex4-1}
\usepackage{amssymb}
\usepackage{amsmath}
\usepackage{graphicx}
\usepackage{amsbsy}
\usepackage{color}
\usepackage{bm}
\newcommand{\bq}{\begin{eqnarray}}
\newcommand{\eq}{\end{eqnarray}}
\newcommand{\bqn}{\begin{eqnarray*}}
\newcommand{\eqn}{\end{eqnarray*}}
\newcommand{\rr}{\mathbf{r}}
\newcommand{\dd}{\mathbf{d}}
\newcommand{\nn}{\mathbf{n}}
\newcommand{\balpha}{{\bar{\alpha}}}
\newcommand{\bbeta}{{\bar{\beta}}}

\begin{document}
%%%%%%%%%%%%%%%%%%%%%%%%%%%%%%%%%%%%%%%%%%%%%%%%%%%%%%%%%%%%%%%%%%%%%%%%%%%%%%
%%%%%%%%%%%%%%%%%%%%%%%%%%%%%%%%%%%%%%%%%%%%%%%%%%%%%%%%%%%%%%%%%%%%%%%%%%%%%%
%%%%%%%%%%%%%%%%%%%%%%%%%%%%%%%%%%%%%%%%%%%%%%%%%%%%%%%%%%%%%%%%%%%%%%%%%%%%%%
\title{Wertheim and Bjerrum-Tani-Henderson theories for associating fluids: a critical assessment}

\author{Riccardo Fantoni}
\email{rfantoni@ts.infn.it}
\affiliation{Dipartimento di Scienze Molecolari e Nanosistemi,
  Universit\`a Ca' Foscari Venezia, Calle Larga S. Marta DD2137,
  I-30123 Venezia, Italy} 

\author{Giorgio Pastore}
\email{pastore@ts.infn.it}
\affiliation{Dipartimento di Fisica dell' Universit\`a and IOM-CNR,
Strada Costiera 11, 34151 Trieste, Italy} 

\date{\today}


\begin{abstract}
Two theories for associating fluids recently used to study 
clustering in models for self-assembling patchy particles,
Wertheim's and Bjerrum-Tani-Henderson theories, are carefully
compared. We show that, for a fluid allowing only for dimerization,
Wertheim theory is equivalent to the Bjerrum-Tani-Henderson theory
neglecting intercluster correlations. Nonetheless, while the former
theory is able to account for percolation and condensation, the latter
is not. For the Bjerrum-Tani-Henderson theory we also rigorously prove
the uniqueness of the solution for the cluster's concentrations and
the reduction of the system of equations to a single one for a single
unknown. We carry out Monte Carlo simulations of two simple models of
dimerizing fluids and compare quantitatively the predictions of the
two theories with the simulation data.  
\end{abstract}

\maketitle
%%%%%%%%%%%%%%%%%%%%%%%%%%%%%%%%%%%%%%%%%%%%%%%%%%%%%%%%%%%%%%%%%%%%%%%%%%%%%%
\section{Introduction}
%%%%%%%%%%%%%%%%%%%%%%%%%%%%%%%%%%%%%%%%%%%%%%%%%%%%%%%%%%%%%%%%%%%%%%%%%%%%%%
\label{sec:introduction}

Recent advances in the experiments and modeling of patchy colloids \cite{Pawar2010,Bianchi2011}, {\sl i.e.} colloidal particles whose interaction is dominated by the presence of selective, short range interaction sites on their surface, have renewed interest in theories able to describe liquid and vapour phases of associating fluids.

Fluid phase theories able to cope with the strong attractions of associating fluids have been developed starting from the seventies, when hydrogen bond in molecular liquids was a prototype problem.
Two of the approximations developed a few decades ago, namely the
approach developed by Tani and Henderson \cite{Tani83}, extending
Bjerrum's theory \cite{Bjerrum1926} for electrolytic solutions, and
the more ambitious statistical mechanics approach by
Wertheim\cite{Wertheim1,*Wertheim2,*Wertheim3,*Wertheim4} have been
recently applied to the study of simple models of patchy colloids \cite{Bianchi2006,Sciortino2007,Bianchi2008,Liu2009,Tavares2010,Russo2011a,Russo2011b,Tavares2012,Rovigatti2013,Fantoni2011,Fantoni2012,Fantoni-Springer2013,Fantoni2013a,Fantoni2013b}.
The novelty introduced by applications to self-assembling colloids is the huge variety of interactions which can be engineered and consequently the richness of the behaviors as far as the cluster population and its dependence on the thermodynamic state are concerned.
Both approaches identify in the fluid and predict populations of suitably defined clusters.

In  both theories, a cluster is defined on the basis of bonding  in configuration space. For example, if we describe
the fluid, as made by particles interacting with a certain
pair-potential $\phi(12)$ between particles 1 and 2, we may consider
two particles as bonded whenever their pair-potential is less than a
given negative value $-\epsilon_{bond}$. Clusters made of one particle
are called ``monomers'', the ones formed by two particles
``dimers'', the ones formed of three particles ``trimers'', ..., and
the ones formed by a higher but small number of particles ``oligomers''. A
cluster made of a number $i$ of particles can also be denoted as an
$i$-mer. If we measure the concentrations of the $i$-mers in an
associating fluid we will find that these are functions of the
thermodynamic state: The temperature $T$ and the density $\rho$ of the
fluid. One can give various definitions of a cluster
\cite{Lee1973,*Ebeling1980,*Gillan1983,*Caillol1995,*Fisher1993,*Friedman1979}
either of a geometrical nature or of a topological one, depending on
the spatial arrangement of the bonded particles. A more physical approach would 
require to introduce the concept of physical cluster \cite{Hill55,Coniglio1977} but virtually all the existing calculations have been  based on clusters defined in configuration space.

In this work we will compare  Wertheim's theory \cite{Wertheim1,*Wertheim2,*Wertheim3,*Wertheim4} and the  one of
Bjerrum-Tani-Henderson \cite{Bjerrum1926,Tani83}. The former one starts from a thorough theoretical analysis, from which  it is possible to derive a
thermodynamic perturbation theory. Here, 
we will only discuss the first order term. At high
temperature the associating fluid reduce to the ``reference'' fluid
that can also be considered as the one obtained from the associating
fluid sending to zero all attractions. The theory is only applicable
when some 
``steric incompatibility'' conditions are fulfilled by the
associating fluid. The latter starts already by the
description of the associating fluid as a mixture of $n_c$ different
species of oligomers where the numbers $N_i$ of $i$-mers are allowed to
vary subject to the constraint of a fixed total number of
particles. One only assumes that the canonical partition function as a
function of all the $N_i$, the volume and the temperature be
factorisable into the product of $n_c$ intra-cluster partition
functions and an inter-cluster partition function. Moreover the
clusters are assumed to interact weakly each other. 

We will show that for $n_c=2$ Wertheim theory coincides with the
Bjerrum-Tani-Henderson theory when the clusters are described as
an ideal gas. Bjerrum-Tani-Henderson theory, on the other hand, allows
to improve on this first level of approximation since one can always
build better approximations to describe the inter-cluster partition
function. In this work we will only consider the Carnahan-Starling
approximation \cite{Carnahan69}, {\sl i.e.} we approximate intercluster correlations with  effective spherically symmetric ones. On the other hand the simple and
elegant theory of Wertheim is able, unlike the Bjerrum-Tani-Henderson
theory, to describe fluids with percolating ($n_c\to\infty$) clusters. Due to this fact
Wertheim's theory is able to describe in a consistent way the liquid phase while the    
Bjerrum-Tani-Henderson one is not. So, for $n_c$ finite,
Bjerrum-Tani-Henderson theory is expected to be more powerful and
flexible than Wertheim theory since it allows to have more accurate
results and it is not restricted to systems obeying the steric
incompatibility conditions. Instead,  Wertheim's theory is the method of choice whenever a
consistent picture of the phase diagram is required.

We will then present a comparison and a critical assessment of the two theories
by comparison with 
new Monte Carlo simulation results for two model fluids with $n_c=2$: 
a binary
mixture and a one-component system, both particularly suitable 
for comparing theories for association.  In particular we will show an,
apparently unavoidable, subtle short-come that may appear in the
Bjerrum-Tani-Henderson when applied to  
multicomponent fluid mixtures: At high temperatures, when the fluid is
dissociated, in the Bjerrum-Tani-Henderson theory one is left with a
one-component mixture of monomers which may differ strongly from the
original multicomponent mixture.    

The paper is organized as follows: In Section \ref{sec:thermodynamics}
we introduce the thermodynamic quantities we will take in
consideration in the following; in Section \ref{sec:theory} we
describe the two association theories discussing the problem of finite
and infinite clusters (Section \ref{sec:clusters}) and the problem of
one attractive site (Section \ref{sec:M=1}); in Section
\ref{sec:binodal} we introduce the problem of the gas-liquid
coexistence; in Section \ref{sec:microscopic} we comment on the
relevance of the pair-potential microscopic level of description; in
Section \ref{sec:results-w} we summarize some results 
obtained applying Wertheim theory to specific fluids with identical
sites and sites of two different kinds; in Section
\ref{sec:results-w-h} we apply the two theories to two simple
dimerizing associating fluids (a binary mixture (Section
\ref{sec:w-h-1-binary}) and a one-component fluid (Section
\ref{sec:w-h-1-one})) and compare them with our Monte Carlo simulation
results; in Section \ref{sec:nc>2} we consider again the problem of
infinite clusters for the Bjerrum-Tani-Henderson theory; Section
\ref{sec:conclusions} summarizes the main results and contains a few final remarks. 

%%%%%%%%%%%%%%%%%%%%%%%%%%%%%%%%%%%%%%%%%%%%%%%%%%%%%%%%%%%%%%%%%%%%%%%%%%%%%%
\section{Thermodynamics}
%%%%%%%%%%%%%%%%%%%%%%%%%%%%%%%%%%%%%%%%%%%%%%%%%%%%%%%%%%%%%%%%%%%%%%%%%%%%%%
\label{sec:thermodynamics}

Consider a one-component fluid of $N$ associating 
particles in a volume $V$ at an absolute temperature $T=1/\beta k_B$
with $k_B$ Boltzmann constant. The inter-particle interaction is assumed to include a hard sphere (HS) part, an isotropic attraction and localized bonding interaction, in general anisotropic.

The Helmholtz free energy $A$ of a hard-sphere associating fluid
can be written as a sum of separate contributions \cite{Jackson88} 
\bq \label{HS-mf-bond}
A=A_{HS}+A_{bond},
\eq 
where $A_{HS}$ is the free energy due to the hard-sphere repulsive
cores and $A_{bond}$ is the change in the free energy due to the
bonding interaction responsible for association.
We will generally use the notation $a(\rho,T)=a=A/N$ for the free
energy per particle, where $\rho=N/V$ is the density of the fluid.

The excess hard-sphere free energy per particle can be modeled by the
Carnahan and Starling \cite{Carnahan69}
\bq \label{Carnahan-Starling}
\beta a_{HS}^{ex}=\frac{4\eta-3\eta^2}{(1-\eta)^2},
\eq
where $\eta=(\pi/6)\rho\sigma^3$ is the packing fraction of the
hard-spheres of diameter $\sigma$. So that adding the ideal gas
contribution $\beta a_{id}=\ln(\rho\Lambda^3/e)$, with $\Lambda$ the
de Broglie thermal wavelength, we obtain $a_{HS}=a_{id}+a_{HS}^{ex}$.

We  can always define a unit of length, ${\cal S}$, and a unit of
energy, ${\cal E}$, so that we can introduce a reduced density,
$\rho^*=\rho{\cal S}^3$, and a reduced temperature, 
$T^*=k_BT/{\cal E}$.

The association contribution $A_{bond}$ will be discussed in the next
section. 

%%%%%%%%%%%%%%%%%%%%%%%%%%%%%%%%%%%%%%%%%%%%%%%%%%%%%%%%%%%%%%%%%%%%%%%%%%%%%%
\section{Bjerrum-Tani-Henderson vs Wertheim}
%%%%%%%%%%%%%%%%%%%%%%%%%%%%%%%%%%%%%%%%%%%%%%%%%%%%%%%%%%%%%%%%%%%%%%%%%%%%%%
\label{sec:theory}

We present now the two association theories of Bjerrum-Tani-Henderson (BTH)
\cite{Tani83} and of Wertheim (W)
\cite{Wertheim1,*Wertheim2,*Wertheim3,*Wertheim4}. We derive in each
case the bond free energy per particle $a_{bond}$ such that the full
free energy per particle of the associating fluid can be written as
$a=a_0+a_{bond}$, where $a_0=a_{id}+a^{ex}_0$ is the contribution of
the reference fluid, the one obtained from the associating fluid
setting to zero all the bonding localized attractions. 

%%%%%%%%%%%%%%%%%%%%%%%%%%%%%%%%%%%%%%%%%%%%%%%%%%%%%%%%%%%%%%%%%%%%%%%%%%%%%%
\subsection{Bjerrum-Tani-Henderson thermodynamic theory}
%%%%%%%%%%%%%%%%%%%%%%%%%%%%%%%%%%%%%%%%%%%%%%%%%%%%%%%%%%%%%%%%%%%%%%%%%%%%%%
\label{sec:henderson}

We assume that our fluid is composed of $n_c$ species of clusters. The
species $i$ contains $N_i$ clusters each made of $i$ particles. Tani and
Henderson
\cite{Tani83,Fantoni2011,Fantoni2012,Fantoni-Springer2013,Fantoni2013a,Fantoni2013b} 
assumed that the total partition function of the fluid can be written
factorizing the $n_c$ intra-cluster partition functions of the single
clusters known a priori as functions of the temperature $T$
alone. Moreover, assuming that the inter-cluster partition function can
be approximated treating the (weakly interacting) clusters as
hard-spheres of diameter $\sigma_c$, they find the following solution
as a result of an extremum procedure 
\bq \label{N1}
N_1&=&N\lambda z_1/\rho G(\eta_c),\\ \label{Ni}
N_i&=&N_1\lambda^{i-1}z_i/z_1,~~~i=1,2,\ldots,n_c
\eq
with
\bq \label{constraint}
N&=&\sum_{i=1}^{n_c} i N_i,\\ 
N_c&=&\sum_{i=1}^{n_c} N_i < N,
\eq
where $N$ is the total number of particles, $\rho=N/V$ is the density
of the fluid, $N_c$ the total number of clusters, $\rho_c=N_c/V$ is
the density of the clusters, $\eta_c=(\pi/6)\rho_c\sigma_c^3$ is the
packing fraction of the clusters of diameter $\sigma_c$, $z_i>0$ the
intra-cluster configuration partition function for the species $i$
($z_1=1$ by definition), and $\lambda>0$ is determined through the
constraint of Eq. (\ref{constraint})
\bq \label{lambda}
0&=&\sum_{i=1}^{n_c} i\lambda^i z_i-\rho G(\eta_c),\\ \label{G}
G(x)&=&\exp\left[\frac{d(x\beta a^{ex}_0(x))}{dx}\right]
=\exp\left[\frac{x(8-9x+3x^2)}{(1-x)^3}\right],
\eq
where $a^{ex}_0(\eta)=a^{ex}_{HS}$. This equation for the unknown parameter $\lambda$ always admits a unique solution. In fact, $G(x)$ is a strictly monotonous
increasing function of $0\le x <1$ with $G(0)=1$ and $\lim_{x\to
  1^-}G(x)=+\infty$. We introduce the concentration of clusters of
species $i$, the $i$-mers, as $x_i=N_i/N$, and the total concentration
of clusters $x_c=N_c/N=\sum_{i=1}^{n_c}x_i=\sum_{i=1}^{n_c}\lambda^iz_i/
\sum_{i=1}^{n_c}i\lambda^iz_i$. Then we notice that $\lim_{\lambda\to
  0}x_c=1$, $\lim_{\lambda\to\infty}x_c=1/n_c<1$, and $x_c$ is a
strictly monotonous decreasing function of $\lambda$ \cite{note0}. So
$G(\eta_c)$ 
is a strictly monotonous decreasing function of $\lambda$ with
$\lim_{\lambda\to 0}G(\eta_c)=G[(\pi/6)\rho\sigma_c^3]$ and
$\lim_{\lambda\to\infty}G(\eta_c)=G[(\pi/6n_c)\rho\sigma_c^3]$. We
also notice that we must require $(\pi/6)\rho\sigma_c^3<1$. Observing
next that $\sum_{i=1}^{n_c} i\lambda^i z_i$ is a strictly monotonous
increasing function of $\lambda$ which is zero at $\lambda=0$, we
conclude that Eq. (\ref{lambda}) must admit always only one solution
$\lambda>0$ such that $\lim_{\rho\to 0}\lambda=0$ and
$\lim_{\rho\to 0}x_1=1$.  

The total partition function $Q_{tot}$ of the fluid is given then by 
\bq \nonumber
\ln Q_{tot} &=& \sum_i[N_i\ln z_i-(N_i\ln N_i-N_i)]+\ln Z_c\\
&=&N_c-N_c\ln N_1-(N-N_c)\ln\lambda+\ln Z_c,
\eq
where $Z_c$ is the inter-cluster configurational partition function and
$\beta A_c^{ex}=-\ln(Z_c/V^{N_c})$ is the inter-cluster excess free energy.

Introducing the concentration of monomers $x_1=N_1/N$ and the
concentration of clusters $x_1<x_c=N_c/N<1$ (note that $1/x_c$ can be
considered as a measure of the average cluster size) we can rewrite
\bq \nonumber
\beta a_{bond}^{BTH}&=&\beta\left[a-\left(a^{id}+a_0^{ex}\right)\right]
\\ \label{bond-h}
&=&x_c\ln x_1+(1-x_c)\ln(\lambda
e/\rho)+\beta\left(a_c^{ex}-a_0^{ex}\right)+constants,  
\eq
where $\beta a=-(\ln Q_{tot})/N$ is the associating fluid total free
energy per particle and $a_0^{ex}+a^{id}$ is the reference system
total free energy per particle. Note that, in the absence of
attractions and therefore in the presence of monomers only $x_1=x_c=1$,
in order to have $a_{bond}^{BTH}=0$ we must have
$a_0^{ex}=\lim_{x_c\to 1}a_c^{ex}$. Only for $\sigma_c=\sigma$ this
condition is satisfied by the Carnahan-Starling reference system,
$a_{HS}^{ex}$ of Eq. (\ref{Carnahan-Starling}). In the most general
case we may think at $\sigma_c$ as a function of the thermodynamic
state of the associating fluid. In the present work we will always
restrict to the case of a constant $\sigma_c$.

At high temperatures all $z_i\to 0$ for $i>1$ and $x_1\to x_c\to 1$ or
$\lambda\to \rho G[(\pi/6)\rho\sigma_c^3]/z_1$,
which means we have complete dissociation. At low temperatures
all $z_i\to\infty$ for $i>1$ and $x_1\to 0$ or $\lambda\to 0$, which
means that we have association. 

%%%%%%%%%%%%%%%%%%%%%%%%%%%%%%%%%%%%%%%%%%%%%%%%%%%%%%%%%%%%%%%%%%%%%%%%%%%%%%
\subsection{Wertheim thermodynamic theory}
%%%%%%%%%%%%%%%%%%%%%%%%%%%%%%%%%%%%%%%%%%%%%%%%%%%%%%%%%%%%%%%%%%%%%%%%%%%%%%
\label{sec:wertheim}

In  Wertheim theory \cite{Wertheim1,*Wertheim2,*Wertheim3,*Wertheim4}
one assumes that each hard-sphere of the one-component fluid (the case
of a mixture will be considered in detail in Section
\ref{sec:results-w-h-1}) is decorated with a set $\Gamma$ of $M$
attractive sites. 
Under the assumptions of: [i.] a single bond per site, [ii.] no
more than one bond between any two particles, and [iii.] no closed
loop of bonds, one can write in a first order thermodynamic
perturbation theory framework, valid at reasonably high temperatures, 
\bq \label{bond-w}
\beta a_{bond}^{W}=\sum_{\alpha\in\Gamma}\left(\ln x_\alpha -
\frac{x_\alpha}{2}\right) + \frac{M}{2},
\eq 
where $x_\alpha=N_\alpha/N$ is the fraction of sites $\alpha$ that are
not bonded (not to be confused with $x_i$ the concentration of clusters
made of a number $i$ of particles. We will always use a Greek index to
denote a specific site) and can be solved by the ``law of mass action'' 
\bq \label{x-wertheim}
x_\alpha=\frac{1}{1+\rho\sum_{\beta\in\Gamma}x_\beta\Delta_{\alpha\beta}},
~~~\alpha\in\Gamma
\eq
where the probability to form a bond, once the available sites of the
two particles are chosen, is given by
$\rho\Delta_{\alpha\beta}=\rho\Delta_{\beta\alpha}$ and approximated as
\bq \label{Delta-w}
\Delta_{\alpha\beta}=\int_{v_{\alpha\beta}}g_0(r_{12})\langle
f_{\alpha\beta}(12)\rangle_{\Omega_1,\Omega_2} d\rr_{12}. 
\eq
Here the integral is over the volume $v_{\alpha\beta}$ of the bond
$\alpha\beta$, $g_0$ is the radial distribution function of the
reference system, $f_{\alpha\beta}$ is the Mayer function between
site $\alpha$ on particle 1 and site $\beta$ on particle 2 (see
Section \ref{sec:microscopic}), and
$\langle \ldots \rangle_{\Omega_1,\Omega_2}$ denotes an angular
average over all orientations of particles 1 and 2 at a fixed relative
distance $r_{12}$. Eq. (\ref{x-wertheim}) should be solved for the
real physically relevant solution such that $\lim_{\rho\to 0}
x_\alpha=1$.  

At high temperatures $\Delta_{\alpha\beta}\to 0$ and $x_\alpha\to 1$,
which means we have complete dissociation. At low temperatures
(Wertheim theory is a high temperature expansion but here we just
mean the formal low $T$ limit of the first order Wertheim results)
$\Delta_{\alpha\beta}\to\infty$ and $x_\alpha\to 0$, which means that
we have complete association. 

The number of attractive sites controls the physical behavior. Models
with one site allow only dimerization. The presence of two sites
permits the formation of chain and ring polymers. Additional sites
allow formation of branched polymers and amorphous systems.

%%%%%%%%%%%%%%%%%%%%%%%%%%%%%%%%%%%%%%%%%%%%%%%%%%%%%%%%%%%%%%%%%%%%%%%%%%%%%%
\subsubsection{Finite vs infinite clusters}
%%%%%%%%%%%%%%%%%%%%%%%%%%%%%%%%%%%%%%%%%%%%%%%%%%%%%%%%%%%%%%%%%%%%%%%%%%%%%%
\label{sec:clusters}

Wertheim theory, unlike BTH one, allows for the existence 
of infinite clusters in the fluid: The percolation phenomenon. In particular, in
Wertheim theory one can define \cite{Tavares2010} $P_s=\sum_i ix_i$ as
the probability to have a particle in a finite cluster (in BTH theory
$P_s=1$ by construction). One can then define the mean cluster size,
or number averaged size of the finite clusters, $N_n=\sum_i
ix_i/\sum_i x_i$, the mean size of a cluster to which a randomly
chosen particle belongs, or weight averaged cluster size, $N_w=\sum_i
i^2x_i/\sum_i ix_i$, or higher moments of the cluster size
distribution $x_i$.  

The interplay between condensation and clustering in associating
fluids has been the subject of many studies \cite{Tavares2010}. In
particular, Coniglio et al. \cite{Coniglio1977} proposed a general
theory of the equilibrium distribution of clusters, establishing a
relation between percolation and condensation. Percolation is
generally believed to be a prerequisite for condensation. As a matter
of fact in Section \ref{sec:nc>2} we will show explicitly that BTH
theory is unable to account for condensation.

%%%%%%%%%%%%%%%%%%%%%%%%%%%%%%%%%%%%%%%%%%%%%%%%%%%%%%%%%%%%%%%%%%%%%%%%%%%%%%
\subsubsection{One attractive site}
%%%%%%%%%%%%%%%%%%%%%%%%%%%%%%%%%%%%%%%%%%%%%%%%%%%%%%%%%%%%%%%%%%%%%%%%%%%%%%
\label{sec:M=1}

The simplest case we can consider in Wertheim theory is the one with a
single site $\alpha$, $M=1$. In this case only monomers and dimers can ever
form. Solving the law of mass action for $x=x_\alpha$, the fraction of
non-bonded sites $\alpha$ which coincides with the concentration of
monomers $x_1$, we find 
\bq
x=\frac{2}{1+\sqrt{1+4\rho\Delta}},
\eq
with $\Delta=\Delta_{AA}$. Which has the correct low density limit
$\lim_{\rho\to 0}x=1$.

Analogously we can solve this simple case in BTH theory allowing
only for monomers and dimers, $n_c=2$, and choosing the ideal gas
approximation for the inter-cluster configurational partition
function, $G=1$ (the $\sigma_c\to 0$ limit of Eq. (\ref{G})). Then we
should solve for $\lambda>0$ in the following quadratic equation
\bq \label{one-x1}
x_1&=&\lambda z_1/\rho,\\
x_2&=&\lambda^2 z_2/\rho,\\
1&=&x_1+2x_2.
\eq
The solution for the monomers concentration is 
\bq
x_1=\frac{2}{1+\sqrt{1+8\rho z_2/z_1^2}}.
\eq
We then see that we have agreement between the two theories if we
choose
\bq \label{w-bth-1}
\Delta=2z_2/z_1^2=2z_2.
\eq

Already for this simple case we see that the bond contribution to the
free energy predicted by the two theories, Eq. (\ref{bond-w}) and
Eq. (\ref{bond-h}), coincide. In fact, from BTH theory of
Eq. (\ref{bond-h}), since the excess free energy of the reference
system and the inter-cluster excess free energy are both zero, we
find, up to an additive constant,
\bq \nonumber
\beta a_{bond}^{BTH}&=&x_c\ln x_1+(1-x_c)\ln(\lambda e/\rho)\\ \nonumber
&=&\ln x_1+(1-x_c)\\
&=&\ln x_1 -x_1/2 +1/2=\beta a_{bond}^{W},
\eq 
where the second equality follows from Eq. (\ref{one-x1}), the third
one from observing that $x_2=(1-x_1)/2$, and the last one from
Eq. (\ref{bond-w}).  

BTH theory, on the other hand, allows to be more accurate and to
treat the fluid of clusters instead of just as an ideal gas as a fluid
of hard-spheres of diameter $\sigma_c$. In this case one should solve
numerically Eqs. (\ref{N1}), (\ref{Ni}), and (\ref{lambda}) with $G$
given by Eq. (\ref{G}).  
And the inter-cluster excess free energy will be given by
\bq
\beta a_c^{ex}=\frac{4\eta_c-3\eta_c^2}{(1-\eta_c)^2},
\eq
whereas the excess free energy per particle of the reference system
will be the usual Carnahan-Starling one of
Eq. (\ref{Carnahan-Starling}) \cite{note1}. 

\begin{figure}[htbp]
\begin{center}
\includegraphics[width=10cm]{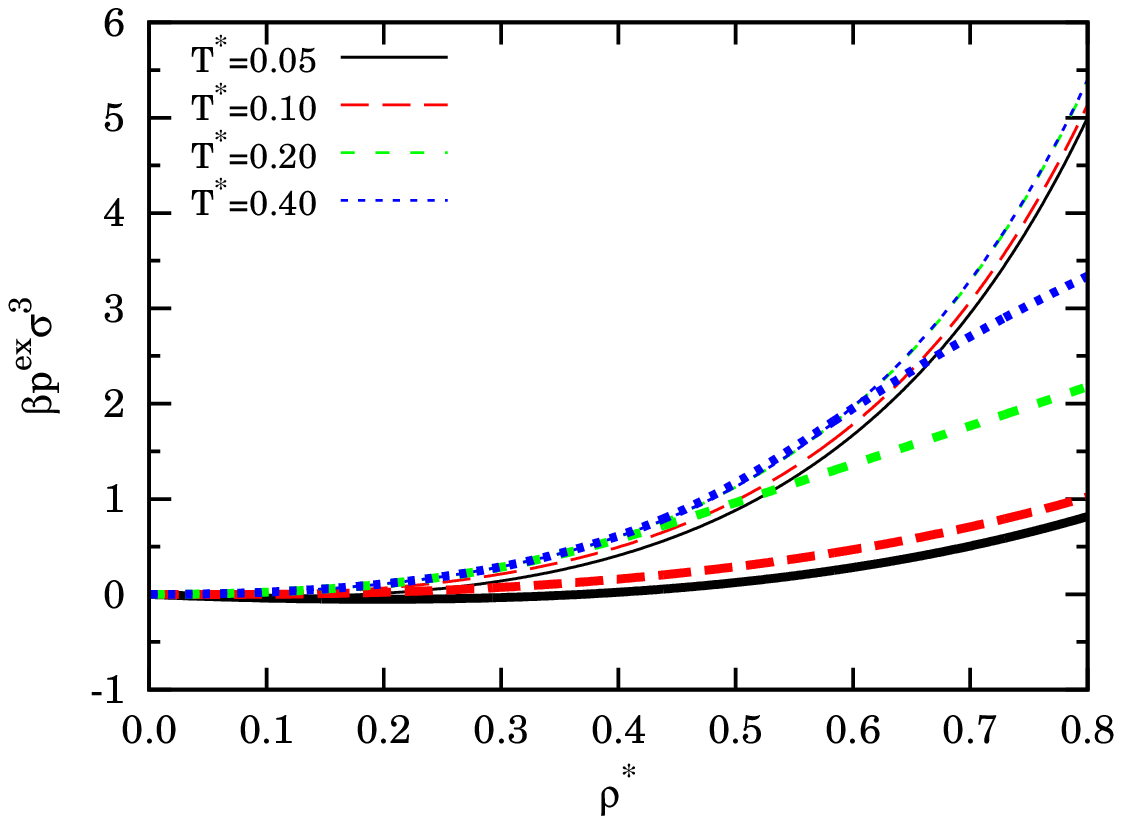}\\
\includegraphics[width=10cm]{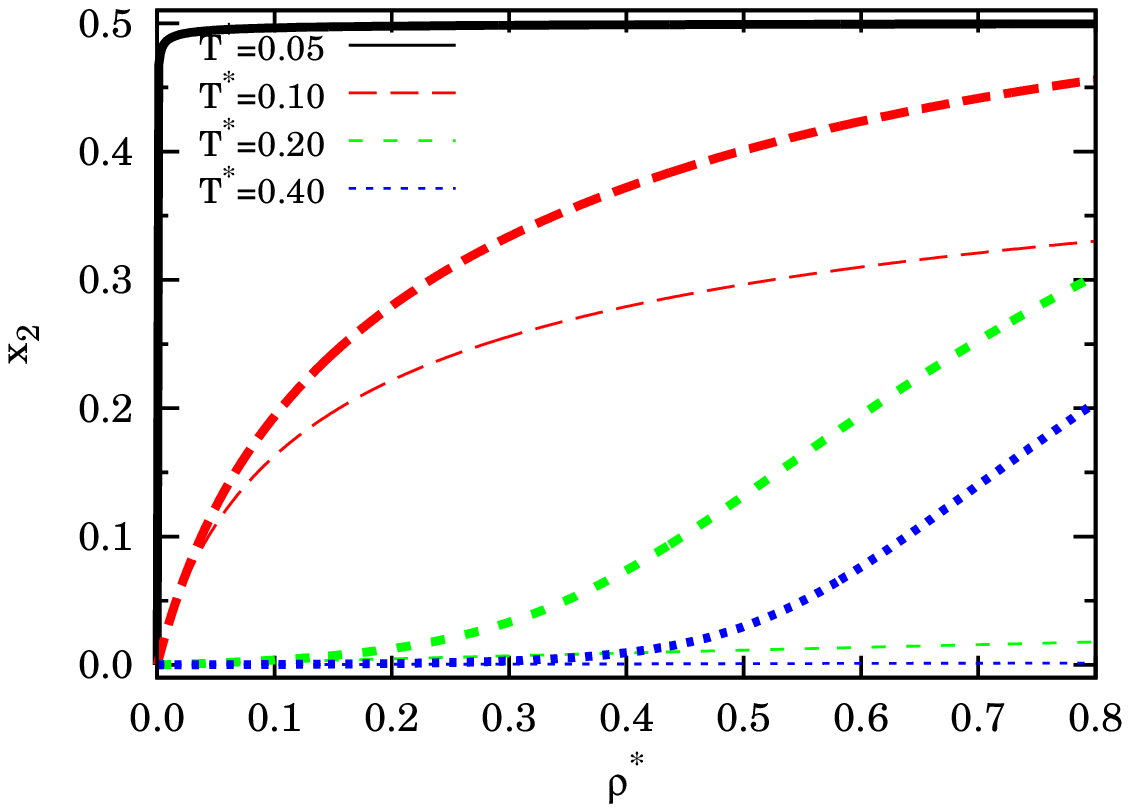}\\
\end{center}  
\caption{(color online) Comparison of the behavior of the excess
  pressure, $\beta p^{ex}=\beta p-\rho$, (top panel) and dimers
  concentration (bottom panel) as functions of density for
  the BTH theory (thick lines), for $\sigma_c=\sigma$, and the W
  theory (thin lines), on several isotherms.}
\label{fig:pr_wt-1}
\end{figure}

Taking $a=a_{HS}+a_{bond}$ and choosing $z_2=\Delta/2$ we compared the
behavior of the two theories. Following
Ref. \onlinecite{Sciortino2007} and approximating the radial
distribution function of the reference system, in Eq. (\ref{Delta})
which appears next in the text, with its zero density  
limit, we choose $\Delta=K^0[\exp(\beta\epsilon)-1]$ with
$K^0=\pi d^4(15\sigma+4d)/30\sigma^2\approx 0.332\times
10^{-3}\sigma^3$. This choice is dictated by the fact that Wertheim
theory gives only a semi-quantitative agreement with simulation data
and we did not find any substantial improvement, at least in the
density ranges we considered, by choosing a better refined low density
approximation, as is done in other works 
\cite{Sciortino2007,Russo2011b}. In Fig. \ref{fig:pr_wt-1} we show 
the comparison of the behavior of the pressure (from Eq. (\ref{z})
which appears next in the text) 
and dimers concentration as functions of density calculated
analytically in Wertheim theory and numerically in BTH theory with
$\sigma_c=\sigma$,  on several isotherms. As expected even at 
very small temperatures there is no sign of a gas-liquid coexistence,
the pressure being a monotonously increasing function of density. 
We have just shown that at low density the two theories must coincide
since $\lim_{\rho\to 0}G=1$, but from the figure we see that the
interval of densities over which the two theories agree increases of
width as $T$ increases. The figure shows how at high temperatures the
two theories tend to become coincident but at low temperatures they
differ strongly. This raises the question of which one of the two
theories is a better approximation when compared to the exact Monte
Carlo results. We will delay the answer to this legitimate question
until Section \ref{sec:w-h-1-one}. BTH theory naturally demands an
approximation for the intra-cluster partition functions. In this work,
unlike previous ones
\cite{Tani83,Fantoni2011,Fantoni2012,Fantoni-Springer2013,Fantoni2013a,Fantoni2013b},
we will always use the relation (\ref{w-bth-1}) when comparing the two
theories. 

Nonetheless we expect Wertheim theory to become more simple and
elegant than BTH theory for $M>1$. As a matter of fact we expect in
these cases the presence in the fluid of $i$-mers of any size
$i$. So that using BTH theory we will necessarily introduce the
additional approximation of the maximum number of cluster species
$i\le n_c$, an artificial cutoff not needed in Wertheim theory.

%%%%%%%%%%%%%%%%%%%%%%%%%%%%%%%%%%%%%%%%%%%%%%%%%%%%%%%%%%%%%%%%%%%%%%%%%%%%%%
\subsection{The gas-liquid coexistence}
%%%%%%%%%%%%%%%%%%%%%%%%%%%%%%%%%%%%%%%%%%%%%%%%%%%%%%%%%%%%%%%%%%%%%%%%%%%%%%
\label{sec:binodal}

In order to determine the gas-liquid coexistence line (the binodal)
one needs to find the compressibility factor $z=\beta p/\rho$, with $p$
the pressure, and the chemical potential $\mu$ of the associating
fluid according to the thermodynamic relations
\bq \label{z}
z(\rho,T)&=&\rho\left(\frac{\partial\beta
  a}{\partial\rho}\right)_{T,N},\\ \label{bm}
\beta\mu(\rho,T)&=&\left(\frac{\partial\beta
  a\rho}{\partial\rho}\right)_{T,V}=z+\beta a.
\eq

The coexistence line is then given by the Gibbs equilibrium condition
of equality of the pressures and chemical potentials of the two phases
\bq \label{gibbs-1}
\rho_gz(\rho_g,T)&=&\rho_lz(\rho_l,T),\\ \label{gibbs-2}
\beta\mu(\rho_g,T)&=&\beta\mu(\rho_l,T),
\eq
from which one can find the coexistence density of the gas $\rho_g(T)$
and of the liquid $\rho_l(T)$ phases.

The critical point $(\rho_c,T_c)$ is determined by solving the
following system of equations
\bq \label{cp1}
\left.\frac{\partial z\rho}{\partial\rho}\right|_{\rho_c,T_c}=0,\\ \label{cp2}
\left.\frac{\partial^2 z\rho}{\partial\rho^2}\right|_{\rho_c,T_c}=0.
\eq
 
%%%%%%%%%%%%%%%%%%%%%%%%%%%%%%%%%%%%%%%%%%%%%%%%%%%%%%%%%%%%%%%%%%%%%%%%%%%%%%
\subsection{Microscopic description: Importance of the pair potential}
%%%%%%%%%%%%%%%%%%%%%%%%%%%%%%%%%%%%%%%%%%%%%%%%%%%%%%%%%%%%%%%%%%%%%%%%%%%%%%
\label{sec:microscopic}

The fluid is assumed to be made of particles interacting only through
a pair-potential $\phi(12)=\phi(\rr_1,\Omega_1,\rr_2,\Omega_2)$ where
$\rr_i$ and $\Omega_i$ are the position vector of the center of
particle $i$ and the orientation of particle $i$ respectively.

To give structure to the fluid we further assume that the particles
have an isotropic hard-core of diameter $\sigma$ with
\bq \label{wertheim-potential}
\phi(12)=\phi_{HS}(r_{12})+\Phi(12),
\eq
where $r_{12}=|\rr_{12}|=|\rr_2-\rr_1|$ is the separation between the
two particles 1 and 2 and  
\bq
\phi_{HS}(r)=\left\{\begin{array}{ll}
+\infty & r\le\sigma\\
0       & r>\sigma
\end{array}
\right.,
\eq

The anisotropic part $\Phi(12)$ in Wertheim theory is generally chosen
as
\bq \label{site-site-potential}
\Phi(12)=\sum_{\alpha\in\Gamma}\sum_{\beta\in\Gamma}
\psi_{\alpha\beta}(r_{\alpha\beta}), 
\eq
where
\bq
\rr_{\alpha\beta}=\rr_2+\dd_\beta(\Omega_2)-\rr_1-\dd_\alpha(\Omega_1),
\eq
is the vector connecting site $\alpha$ on particle 1 with site $\beta$
on particle 2. Here $\dd_\alpha$ is the vector from the particle
center to site $\alpha$ with $d_\alpha<\sigma/2$. The site-site
interactions $\psi_{\alpha\beta}\le 0$ are assumed to be purely
attractive. The Mayer functions introduced in Section
\ref{sec:wertheim} are then  defined as
$f_{\alpha\beta}(12)=\exp[-\beta\psi_{\alpha\beta}(r_{\alpha\beta})]-1$.

Wertheim theory depends on the specific form of the site-site
potential only through the quantity $\Delta_{\alpha,\beta}$ of
Eq. (\ref{Delta-w}), 
as long as the three conditions of a single bond per site, 
no more than one bond between any two particles, and no closed loop of
bonds, are satisfied. A common choice, for example, is a square-well
form 
\bq \label{site-site-square-well}
\psi_{\alpha\beta}(r)=\left\{\begin{array}{ll}
-\epsilon_{\alpha\beta} & r\le d_{\alpha\beta}\\
0                    & r>   d_{\alpha\beta}
\end{array}
\right.,
\eq
where $\epsilon_{\alpha\beta}>0$ are site-site energy scales, the
wells depths, and $d_{\alpha\beta}$ are the wells widths. In this
case we must have 
$d_\alpha+d_\beta>\sigma-d_{\alpha\beta}$ moreover we will have 
\bq \label{Delta}
\Delta_{\alpha\beta}=K_{\alpha\beta}(\sigma,d_{\alpha\beta},\eta)
(e^{\beta\epsilon_{\alpha\beta}}-1).
\eq
We will also call $\lim_{\rho\to 0}K_{\alpha\beta}=K_{\alpha\beta}^0$
some purely geometric factors. Remember that $\lim_{\rho\to
0}g_0(r)=\Theta(r-\sigma)$ with $\Theta$ the Heaviside 
step function. Another common choice is the Kern-Frenkel patch-patch
pair-potential model \cite{Kern03}.

In BTH theory on the other hand, we are allowed to relax these
conditions and the choice of the pair-potential is more flexible as
long as it includes some attractive component responsible for the
association. 

%%%%%%%%%%%%%%%%%%%%%%%%%%%%%%%%%%%%%%%%%%%%%%%%%%%%%%%%%%%%%%%%%%%%%%%%%%%%%%
\section{Some results from Wertheim theory}
%%%%%%%%%%%%%%%%%%%%%%%%%%%%%%%%%%%%%%%%%%%%%%%%%%%%%%%%%%%%%%%%%%%%%%%%%%%%%%
\label{sec:results-w}

Wertheim theory of associating fluids has been recently  
tested extensively by Sciortino and coworkers.
In a series of papers, they have studied fluids of
hard-spheres with identical sites allowing for ``chaining''
\cite{Bianchi2006,Sciortino2007,Bianchi2008,Liu2009} and with
sites of two different kinds allowing for ``branching''
\cite{Tavares2010,Russo2011a,Russo2011b} and for ``rings'' formation
\cite{Tavares2012,Rovigatti2013}. They showed how the parameter-free
Wertheim theory is flexible enough to accomodate a vast number of
different microscopic pair-potentials descriptions and nonetheless
pointed out some relevant classes of microscopic features giving rise
to specific macroscopic behaviors at the level of the clustering,
the percolation threshold, and the gas-liquid coexistence.

In all these cases $n_c\to\infty$ so they cannot be treated with the
BTH theory which as we will see in Section \ref{sec:nc>2} is unable to
account for the gas-liquid coexistence. 
Thus, in order to compare the two theories we have to choose  different systems.
%%%%%%%%%%%%%%%%%%%%%%%%%%%%%%%%%%%%%%%%%%%%%%%%%%%%%%%%%%%%%%%%%%%%%%%%%%%%%%
\section{Comparison between Wertheim theory and Bjerrum-Tani-Henderson
  theory} 
%%%%%%%%%%%%%%%%%%%%%%%%%%%%%%%%%%%%%%%%%%%%%%%%%%%%%%%%%%%%%%%%%%%%%%%%%%%%%%
\label{sec:results-w-h}

In order to test the accuracy of the Wertheim and BTH theories we
carried out some Monte Carlo (MC) simulations on simple models of associating
fluids. 

%%%%%%%%%%%%%%%%%%%%%%%%%%%%%%%%%%%%%%%%%%%%%%%%%%%%%%%%%%%%%%%%%%%%%%%%%%%%%%
\subsection{One attractive site, $n_c=2$}
%%%%%%%%%%%%%%%%%%%%%%%%%%%%%%%%%%%%%%%%%%%%%%%%%%%%%%%%%%%%%%%%%%%%%%%%%%%%%%
\label{sec:results-w-h-1}

We limit ourselves to the case $n_c=2$ and we
consider two different realizations of this scenario: A binary
mixture and a one-component fluid.

%%%%%%%%%%%%%%%%%%%%%%%%%%%%%%%%%%%%%%%%%%%%%%%%%%%%%%%%%%%%%%%%%%%%%%%%%%%%%%
\subsubsection{A binary mixture}
%%%%%%%%%%%%%%%%%%%%%%%%%%%%%%%%%%%%%%%%%%%%%%%%%%%%%%%%%%%%%%%%%%%%%%%%%%%%%%
\label{sec:w-h-1-binary}

To test the single site case we considered a symmetric binary
mixture of particles with the following pair-potential between a
particle of species $\balpha$ (in this section a Greek index with
an over-bar labels the particle species) and one of species $\bbeta$ a
center-to-center distance $r$ apart 
\bq
\phi_{\balpha\bbeta}(r)=\left\{\begin{array}{ll}
+\infty                        & r\le\sigma_{\balpha\bbeta}\\
-(1-\delta_{\balpha\bbeta})\epsilon  & 
\sigma_{\balpha\bbeta}<r\le \sigma_{\balpha\bbeta}+{\cal W}\\ 
0                               & r> \sigma_{\balpha\bbeta}+{\cal W}
\end{array}
\right.,
\eq
where $\sigma_{\balpha\bbeta}=(1/2)(\sigma_\balpha+\sigma_\bbeta)(1+{\cal
  D}_{\balpha\bbeta})$ with $\sigma_\balpha=\sigma$ and ${\cal
  D}_{\balpha\bbeta}=-(1-\delta_{\balpha\bbeta})$ with $\balpha$ and $\bbeta$ equal to
$1,2$ and $\delta$ the Kronecker delta. So that
$\sigma_{\balpha\bbeta}=\sigma\delta_{\balpha\bbeta}$. $\epsilon>0$ and
${\cal W}>0$ 
are respectively the square well depth and width for the attraction of
unlike particles. Also we choose the symmetric case where the
concentrations of particles of species $\balpha$, ${\cal X}_\balpha=1/2$ for
$\balpha=1,2$. In this case the ideal part of the free energy will be
given by $\beta a_{id}=\ln(\rho\Lambda^3/e)+{\cal X}_1\ln {\cal
  X}_1+{\cal X}_2\ln {\cal X}_2$ where the entropy of mixing, the last
two terms, is just an additive constant.

It is then clear that, for ${\cal W}<\sigma/2$, this model fluid
allows for dimerization only, just as the $M=1$ case of
Wertheim. In fact, whenever two unlike particles bind, a third particle
can never bind to the formed dimer because of the hard-core repulsion
between like particles. Moreover by choosing ${\cal W}$ small at will
we may reach the ideal condition of $\sigma_c=\sigma$ with $\sigma_c$
the diameter of the dimers in the BTH theory. The reference fluid, the
one with $\epsilon=0$, is a symmetric non-additive-hard-sphere (NAHS)
mixture with non-additivity ${\cal D}_{12}=-1$. We will then take
\bq \label{reference-nahsa}
\beta a_0^{ex}=\frac{2\eta-(3/4)\eta^2}{[1-(1/2)\eta]^2}.
\eq

Wertheim theory has been extended to multicomponent mixtures by
Chapman et al. \cite{Chapman1986,*Joslin1987}. For a mixture with a
number $n_s$ of species and $N_\balpha=N{\cal X}_\balpha$ particles of
species $\balpha=1,2,\ldots, n_s$, we have
\bq \label{bond-w-binary}
\beta a_{bond}^W=\sum_{\balpha=1}^{n_s}{\cal X}_\balpha
[\ln x_\balpha-x_\balpha/2+1/2],
\eq 
where $x_\balpha=N_1^\balpha/N_\balpha$ is the monomer fraction of
species $\balpha$, with $N_1^\balpha$ the number of monomers of species
$\balpha$, and is determined by the following law of mass action
\bq \label{x-binary-symmetric}
x_{\balpha}=\frac{1}{1+\rho\sum_{\bbeta=1}^{n_s}
{\cal X}_\bbeta x_\bbeta\Delta_{\balpha\bbeta}},
\eq
where 
\bq
\Delta_{\balpha\bbeta}=\Delta_{\bbeta\balpha}=\int_{v_{\balpha\bbeta}}g^0_{\balpha\bbeta}(r_{12})
\langle f_{\balpha\bbeta}(12)\rangle_{\Omega_1,\Omega_2} d\rr_{12},
\eq
with $g^0_{\balpha\bbeta}$ the partial radial distribution of the
reference fluid and
$f_{\balpha\bbeta}(12)=e^{-\beta[\phi_{\balpha\bbeta}(r_{12})-\phi_{\balpha\bbeta}^0(r_{12})]}-1$
the Mayer function between particle 1 of species $\balpha$ and particle
2 of species $\bbeta$, with $\phi_{\balpha\bbeta}^0$ the pair-potential
of the reference fluid.   

In our symmetric binary case $x_{\balpha=1}=x_{\balpha=2}=x$ and
$\Delta=\Delta_{12}=K_{12}\left(e^{\beta\epsilon}-1\right)$
(with $\Delta_{\balpha\balpha}=0$ for $\balpha=1,2$)  
where, since the unlike radial distribution function of the reference
system is the one of the ideal gas, equal to one everywhere, we have
exactly $K_{12}=(4/3)\pi{\cal W}^3$. The solution of
Eq. (\ref{x-binary-symmetric}) is 
\bq
x=\frac{2}{1+\sqrt{1+2\rho\Delta}}.
\eq
Here we will choose ${\cal W}=0.1\sigma$. 

On the other hand BTH theory continues to hold just as in its one
component fluid formulation given in Section \ref{sec:henderson}. We
expect the cluster diameter to vary within 
the interval $\sigma\le\sigma_c\le\sigma+{\cal W}$ even if for the
comparison with the simulation data we will need to consider
$\sigma_c<\sigma$. We will now choose $z_2=\Delta/4$.

At high temperatures $z_2=\Delta/4\to 0$ and $x_1\to 1, x_c\to 1$
so $\beta a^{W}=\beta a_{id}+[2\eta-(3/4)\eta^2]/[1-(1/2)\eta]^2$
whereas $\beta a^{BTH}=\beta a_{id}+[4\eta_c-3\eta_c^2]/[1-\eta_c]^2$.
Then for $\sigma_c\neq\sigma/2^{1/3}$ the parameter free Wertheim
theory is certainly a better approximation than BTH. 
At low temperatures $z_2=\Delta/4\to\infty$ and $x_1\to 0, x_c\to
1/2$, and the two theories become equivalent for $\sigma_c=\sigma$
(see Appendix \ref{app:1}). Within BTH one is free to choose 
$\sigma_c$ in such way to get more accurate results.

The opposite behavior was observed for the one-component case of
Section \ref{sec:M=1} where the two theories, for $\sigma_c=\sigma$,
become equivalent at high temperature and at low temperature they
differ and BTH is expected to become better than W.

We carried out MC simulations of this mixture in the
canonical ensemble using a total number $N=500$ of particles. In the
simulation we measure the pressure from the virial theorem as
\cite{HanMcD86}  
\bq
z^{MC}=1+\frac{1}{3}\pi\rho\left[\sigma^3g_{11}(\sigma^+)- 
\left(e^{\beta\epsilon}-1\right){\cal W}^3g_{12}({\cal W}^+)\right],
\eq
where $g_{\balpha\bbeta}$ are the partial radial distribution
functions. In the simulation we define a dimer as any two particles
for which the pair-potential equals $-\epsilon$. So, we measure
the dimers concentration $x_2^{MC}=-u^{ex}/\epsilon$, where $u^{ex}$ is the
excess internal energy per particle of the fluid. As usual we choose
$\sigma$ as the unit of length and 
$\epsilon$ as the unit of energy. At the lowest temperature studied,
$T^*=0.1$, the probability of breaking a bond is of the order of
$\exp(1/0.1)$, thus requiring $2\times 10^4$ MC attempts to break such
a bond. Our simulations were of the order of $4\times 10^5$ MC steps
long, with a MC step made by $N$ single particle moves.  

We compare the simulation data with the dimers concentrations, $x_2^W$
and $x_2^{BTH}$, and pressures, $\rho z^W$ and $\rho z^{BTH}$,
predicted by Wertheim and BTH theories, where 
\bq
z^{W}&=&1+\rho\frac{\partial\beta\left(a_0^{ex}+a_{bond}^{W}\right)}
{\partial\rho},\\
z^{BTH}&=&1+\rho\frac{\partial\beta\left(a_0^{ex}+a_{bond}^{BTH}\right)}
{\partial\rho},
\eq  
with $a_0^{ex}$ given by Eq. (\ref{reference-nahsa}),
$a_{bond}^{W}$ given by Eq. (\ref{bond-w-binary}), and
$a_{bond}^{BTH}$ given by Eq. (\ref{bond-h}) with $n_c=2$ 
and $z_2=\Delta/4$. 

\begin{figure}[htbp]
\begin{center}
\includegraphics[width=10cm]{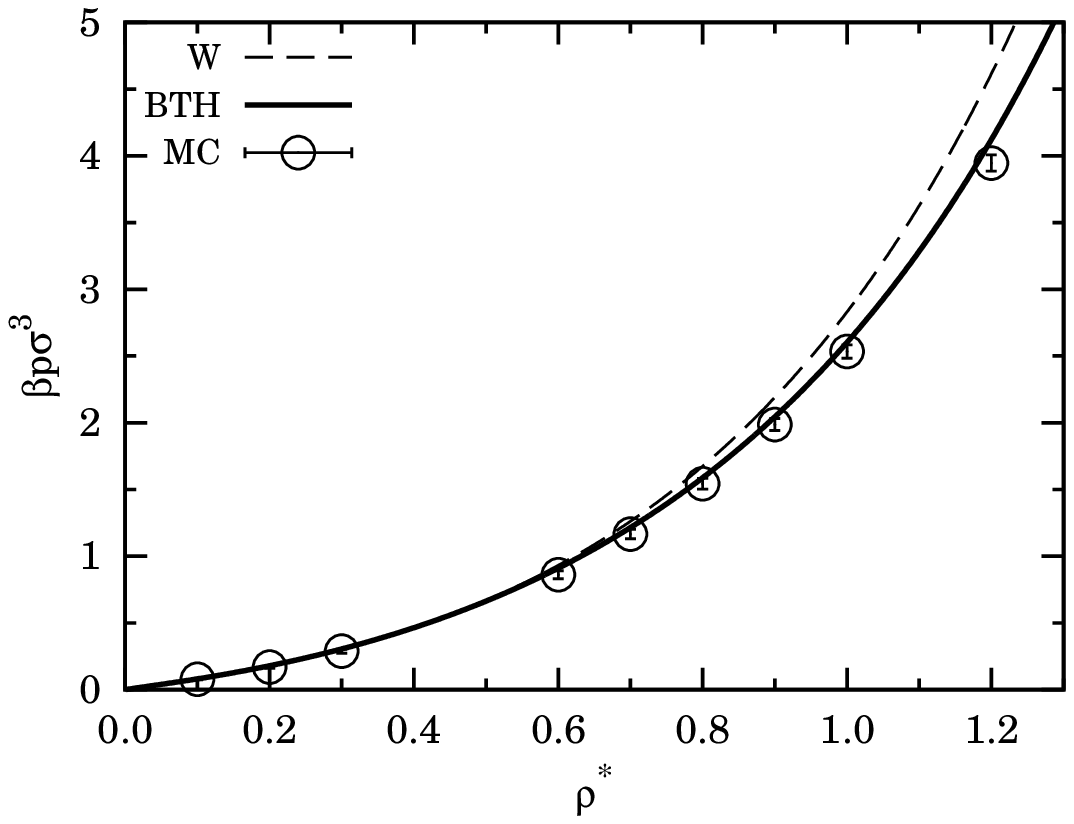}\\
\includegraphics[width=10cm]{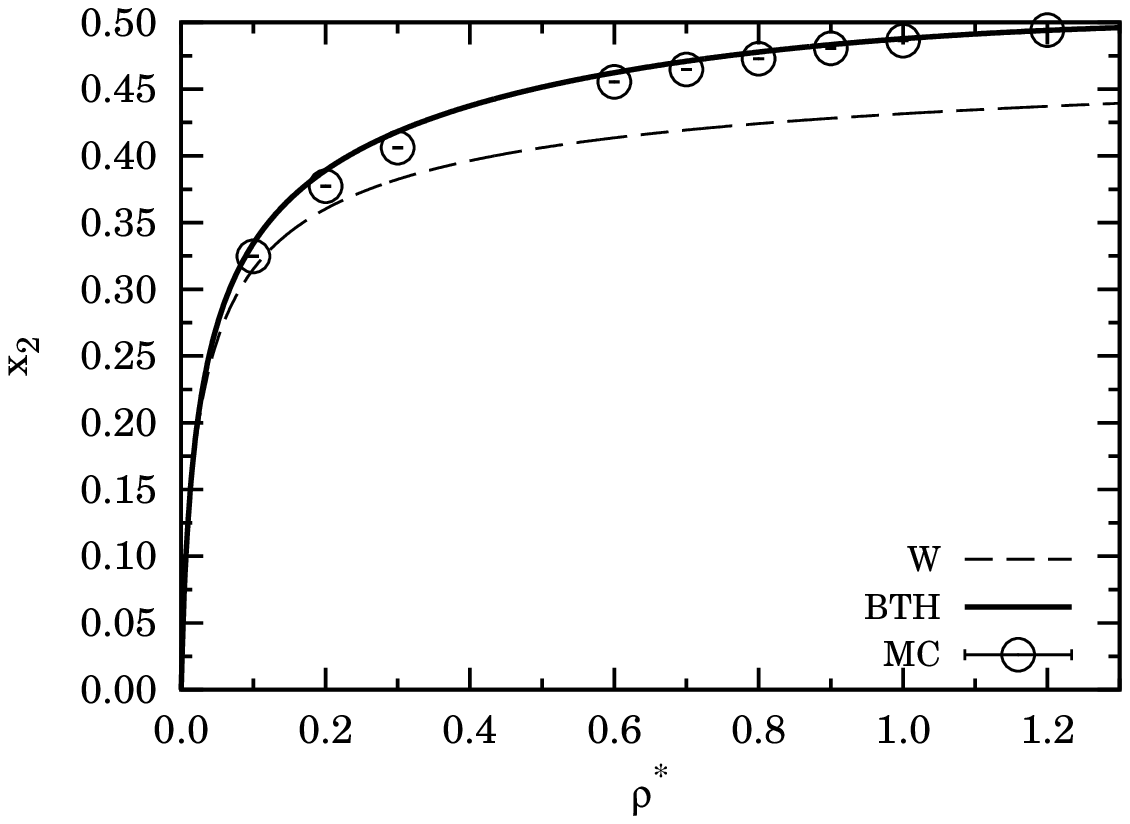}
\end{center}  
\caption{Pressure(top panel) and dimers concentration (bottom panel)
  as a function of density on the $T^*=0.1$ 
  isotherm for ${\cal W}=0.1\sigma$. The broken line is the prediction
  of W theory, the continuous line the one of BTH theory with
  $\sigma_c=0.98\sigma$, and the points are the exact MC data.}
\label{fig:pr_wt-nahsa-t0.1}
\end{figure}

\begin{figure}[htbp]
\begin{center}
\includegraphics[width=10cm]{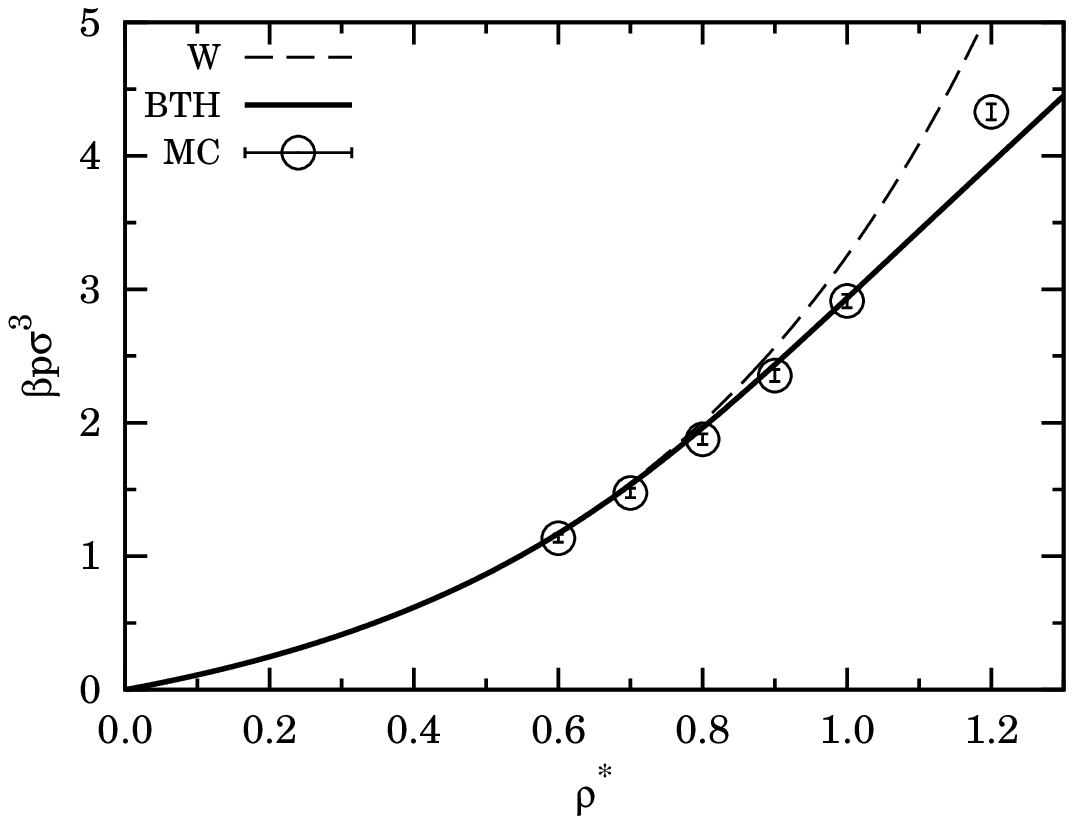}\\
\includegraphics[width=10cm]{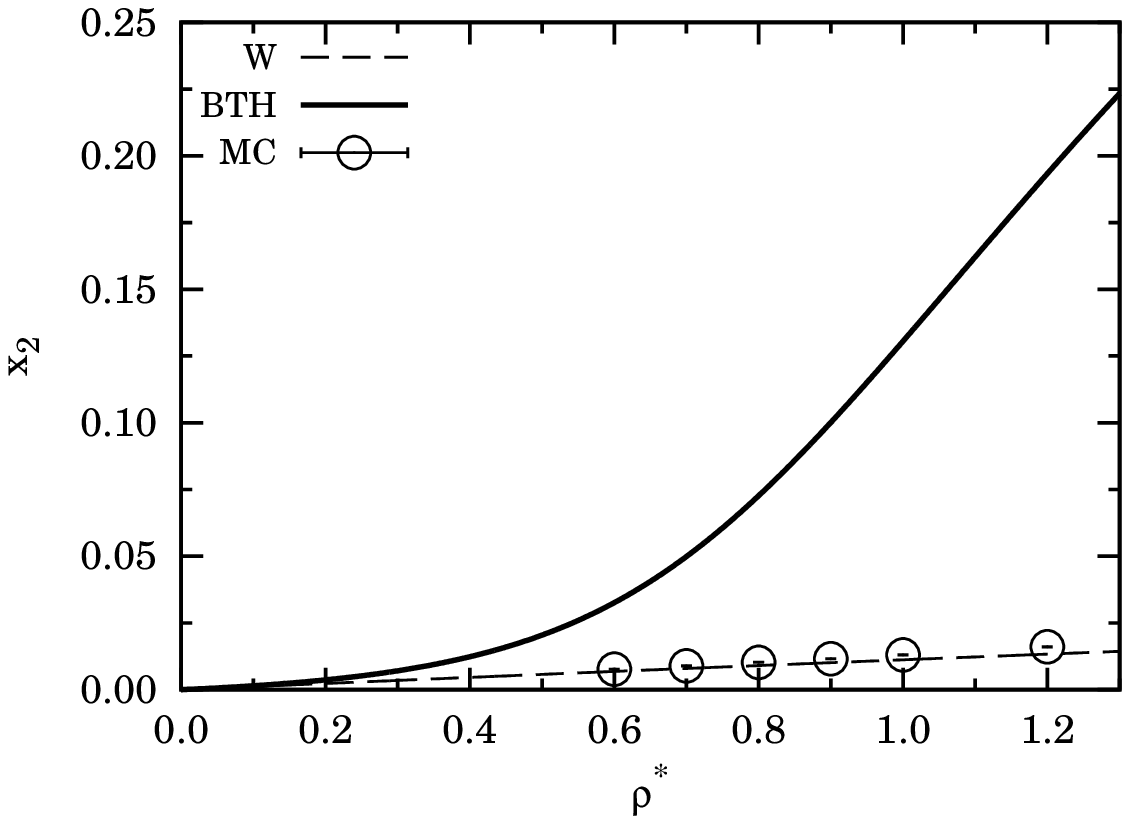}
\end{center}  
\caption{Pressure (top panel) and dimers concentration (bottom panel)
  as a function of density on the $T^*=0.4$ 
  isotherm for ${\cal W}=0.1\sigma$. The broken line is the prediction
  of W theory, the continuous line the one of BTH theory with
  $\sigma_c=\sigma/2^{1/3}$, and the points are the exact MC data.}
\label{fig:pr_wt-nahsa-t0.4}
\end{figure}

In Fig. \ref{fig:pr_wt-nahsa-t0.1} we compare the equation
of state and the dimers concentration as a function of density
predicted by Wertheim and BTH theories with the MC results at 
a low reduced temperature $T^*=0.1$. We see that by choosing the
cluster diameter opportunely, $\sigma_c<\sigma$, one can get the BTH
results for the pressure to overlap with MC data over a wide range of
densities. Fig. \ref{fig:pr_wt-nahsa-t0.4} shows the same comparison 
at the high temperature $T^*=0.4$ for the optimal
$\sigma_c=\sigma/2^{1/3}$. From the 
figures we conclude that BTH theory, with the optimal $\sigma_c$ for
the equation of state, improves at low temperatures,
where it becomes more accurate than Wertheim theory, but fails a
correct descriptions of the clusters concentration at high
temperatures and high densities. By appropriately tuning the cluster
diameter $\sigma_c$ it is possible to get better agreement for the
dimer concentration but then the theory would fail to reproduce the
pressure correctly. So it is never possible to get good agreement for
both the pressure and the dimer concentration. 


\begin{figure}[htbp]
\begin{center}
\includegraphics[width=10cm]{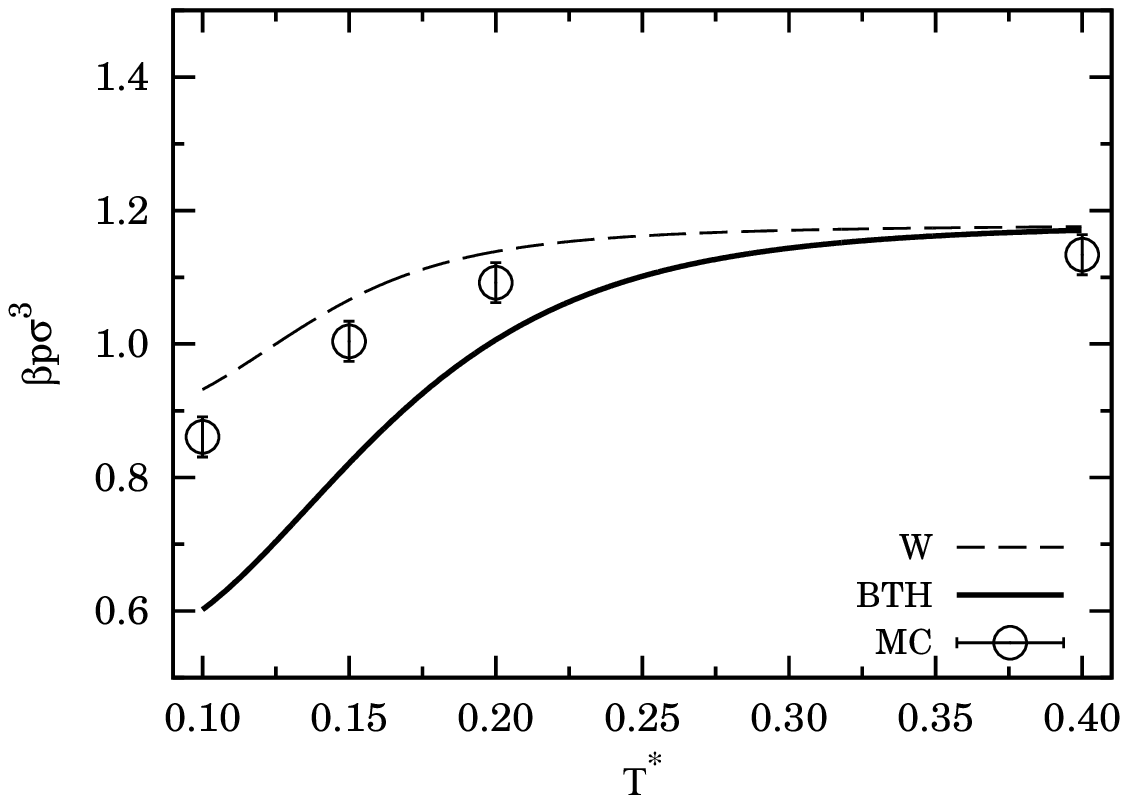}\\
\includegraphics[width=10cm]{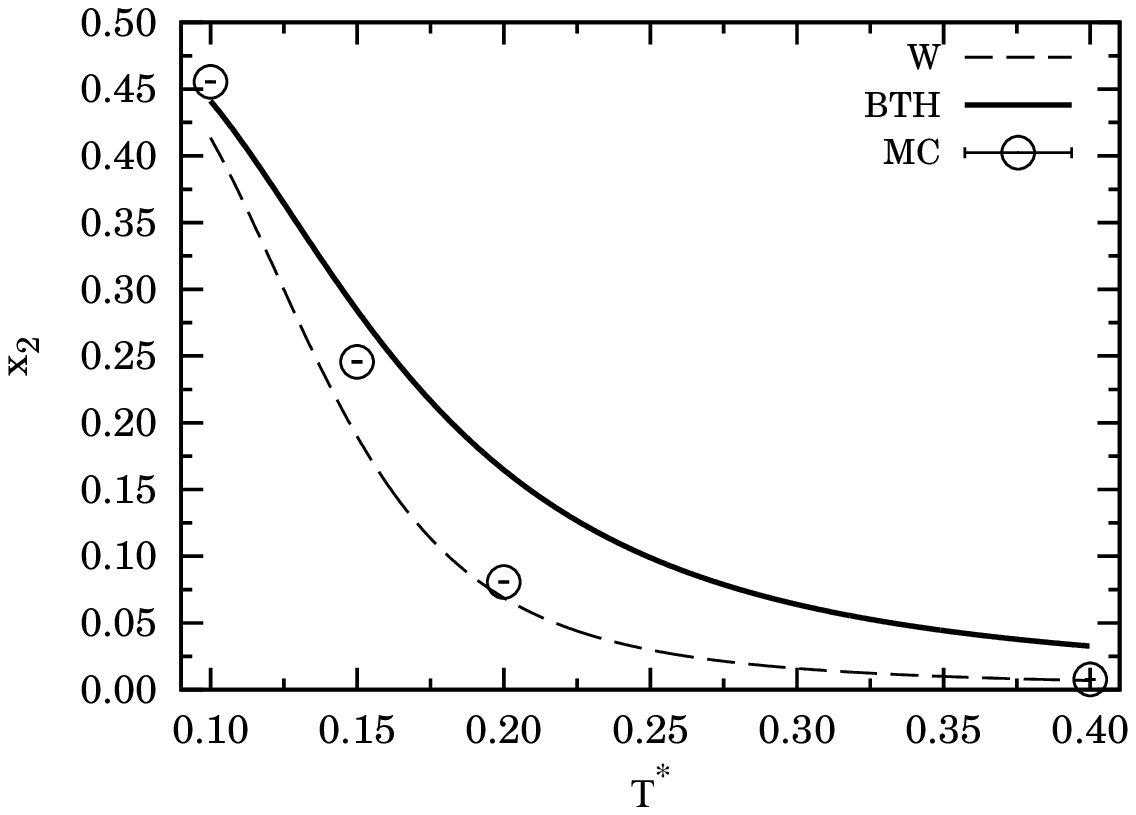}
\end{center}  
\caption{Pressure (top panel) and dimers concentration (bottom panel)
  as a function of temperature on the $\rho^*=0.6$ 
  isochore for ${\cal W}=0.1\sigma$. The broken line is the prediction
  of W theory, the continuous line the one of BTH theory with
  $\sigma_c=\sigma/2^{1/3}$, and the points are the exact MC data.}
\label{fig:pr_wt-nahsa-r0.6}
\end{figure}

In Fig. \ref{fig:pr_wt-nahsa-r0.6} we compare the pressure
and the dimers concentration as functions of temperature
predicted by the two theories, when $\sigma_c=\sigma/2^{1/3}$ in BTH,
with the MC results at a low reduced density $\rho^*=0.6$. The figure
shows how in this case the Wertheim theory is better than BTH.  

%

%%%%%%%%%%%%%%%%%%%%%%%%%%%%%%%%%%%%%%%%%%%%%%%%%%%%%%%%%%%%%%%%%%%%%%%%%%%%%%
\subsubsection{A one-component fluid}
%%%%%%%%%%%%%%%%%%%%%%%%%%%%%%%%%%%%%%%%%%%%%%%%%%%%%%%%%%%%%%%%%%%%%%%%%%%%%%
\label{sec:w-h-1-one}

As a one-component fluid we chose the single patch
Kern-Frenkel model \cite{Kern03,Fantoni2011} where the particles
interact with the following pair-potential
\bq
\phi(r_{12})=\phi_{HS}(r_{12})+\phi_{SW}(r_{12})
\gamma(\hat{\nn}_1,\hat{\nn}_2,\hat{\rr}_{12}),
\eq
where
\bq
\phi_{SW}(r)=\left\{\begin{array}{ll}
-\epsilon & \sigma<r\le\sigma+{\cal W}\\
0         & \mbox{else}
\end{array}
\right.,
\eq
and
\bq
\gamma(\hat{\nn}_1,\hat{\nn}_2,\hat{\rr}_{12})=\left\{\begin{array}{ll}
1 & \hat{\nn}_1\cdot\hat{\rr}_{12}\ge\cos\theta_0~~~\mbox{and}~~~
-\hat{\nn}_2\cdot\hat{\rr}_{12}\ge\cos\theta_0\\
0 & \mbox{else}
\end{array}
\right.,
\eq
here $\hat{\nn}_i$ is a unit vector pointing from the center of particle
$i$ towards the center of her attractive patch and $\theta_0$ is
the angular semi-amplitude of the patch. The fraction of the particle
surface covered by the attractive patch will then be
$\chi=\sqrt{\langle\gamma\rangle_{\Omega_1,\Omega_2}}=\sin^2(\theta_0/2)$. 

In order to have $n_c=2$ we must choose $\theta_0<\pi/6$ or
$\chi<(\sqrt{3}-1)^2/8\approx 0.0670$ in the sticky limit ${\cal W}\to
0$ and
\bq \label{sbpp}
\cos 2\theta_0>\frac{1}{2}+\frac{1}{2\sigma(\sigma+{\cal W})}
\sqrt{6\sigma^3{\cal W}-\sigma^2{\cal W}^2-4\sigma{\cal W}^3-{\cal W}^4},
\eq
more generally, for small ${\cal W}$. As before we choose ${\cal
  W}=0.1\sigma$ and in order to fulfill the single bond per patch
condition (\ref{sbpp}) we take $\theta_0=\pi/12$ or
$\chi=0.0170$. This choice corresponds to a patch-patch bonding volume
$v_{pp}=(\pi/3)[(\sigma+{\cal W})^3-\sigma^3](1-\cos\theta_0)^2\approx
0.402\times 10^{-3}\sigma^3$. We then choose for $\Delta$ its zero
density limit approximation $\Delta=v_{pp}(e^{\beta\epsilon}-1)$.

We carried out MC simulations of this one-component fluid in the
canonical ensemble using a number $N=500$ of particles. The pressure
is calculated during the simulation from the virial 
theorem as follows \cite{Kern03},
\bq
z^{MC}=1+\frac{2\pi}{3}\rho\sigma^3\left[g(\sigma^+)-
(1+{\cal W}/\sigma)^3\left\{g_{pp}\left[(\sigma+{\cal W})^-\right]-
g_{pp}\left[(\sigma+{\cal W})^+\right]\right\}\right],
\eq
where $g_{pp}(r)$ is the radial patch-patch distribution function: The
partial radial distribution function which considers only particles
with facing patches. Again, we measure the dimers concentration as
$x_2^{MC}=-u^{ex}/\epsilon$. As usual we choose $\sigma$ as the unit
of length and $\epsilon$ as the unit of energy. A MC move here
consisted of both a random displacement of the center of the particle
and a random rotation of the particle (according to the Marsaglia
algorithm \cite{Allen}).  

\begin{figure}[htbp]
\begin{center}
\includegraphics[width=10cm]{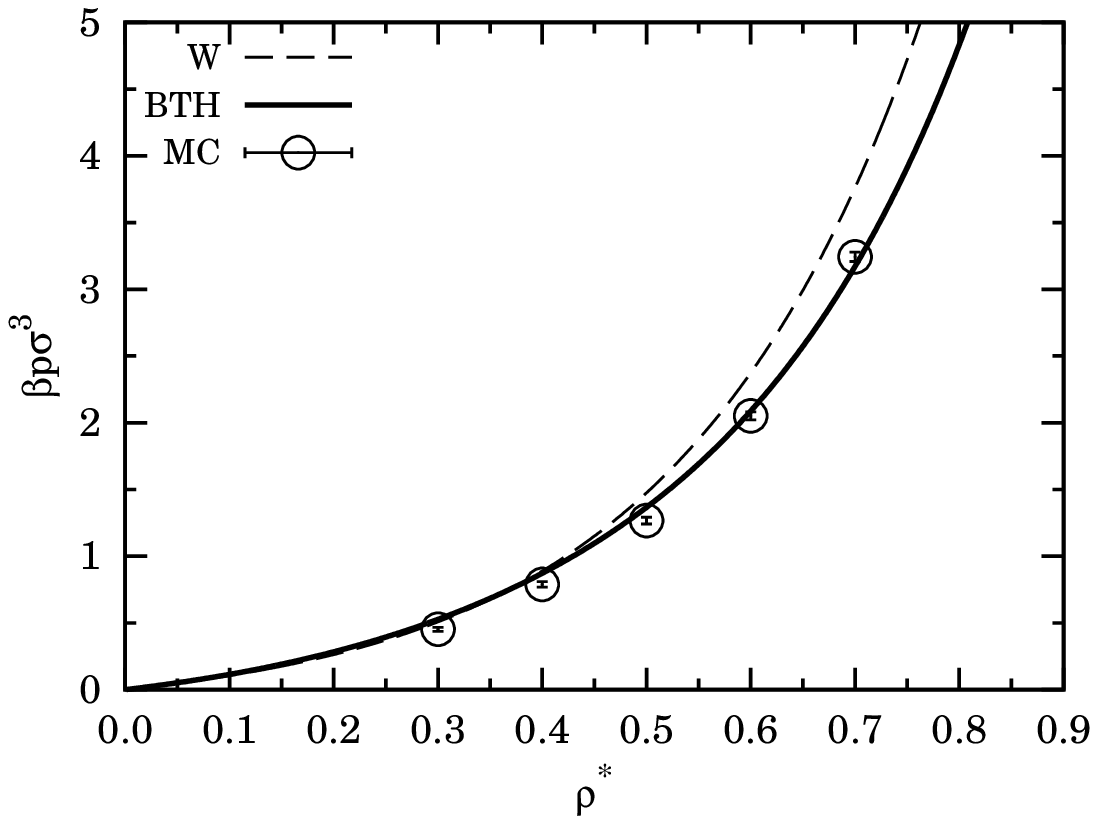}\\
\includegraphics[width=10cm]{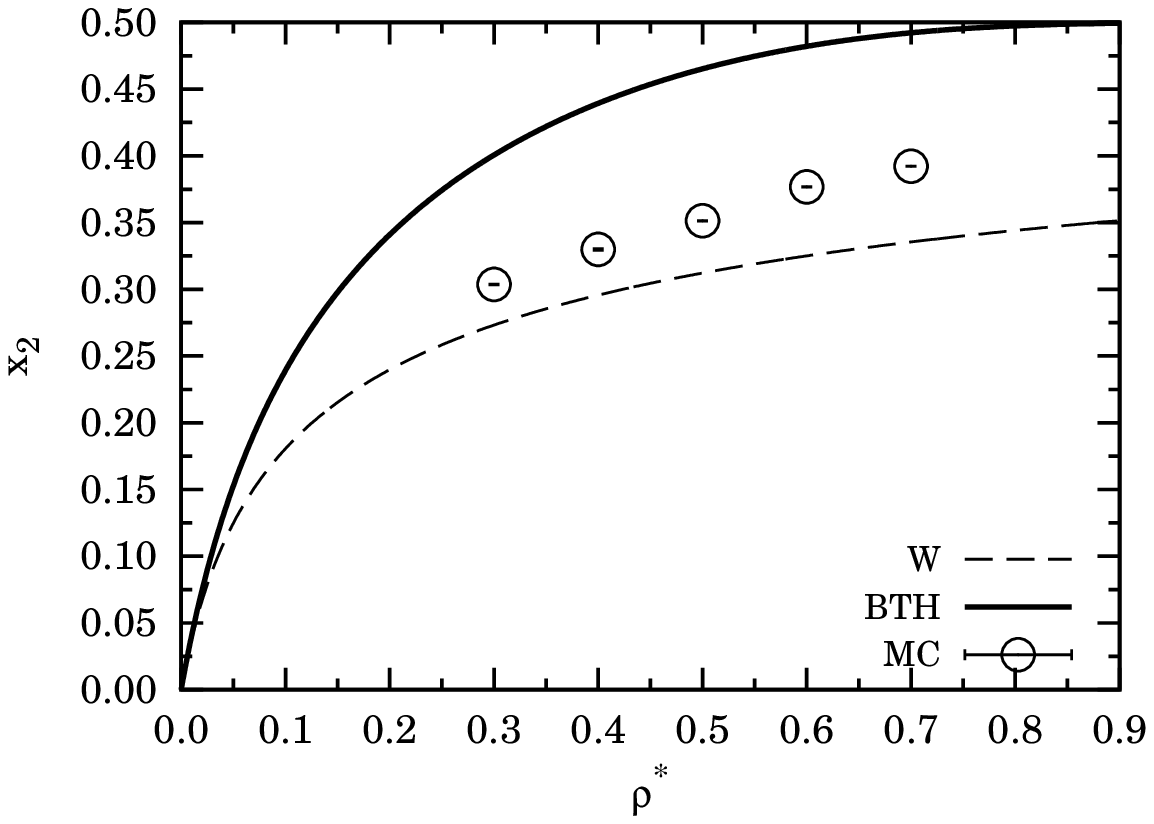}
\end{center}  
\caption{Pressure(top panel) and dimers concentration (bottom panel)
  as a function of density on the $T^*=0.1$ isotherm for ${\cal
    W}=0.1\sigma$ and $\theta_0=\pi/12$. The broken line is the
  prediction 
  of W theory, the continuous line the one of BTH theory with
  $\sigma_c=1.23\sigma$, and the points are the exact MC data.}
\label{fig:pr_wt-kf-t0.1}
\end{figure}

\begin{figure}[htbp]
\begin{center}
\includegraphics[width=10cm]{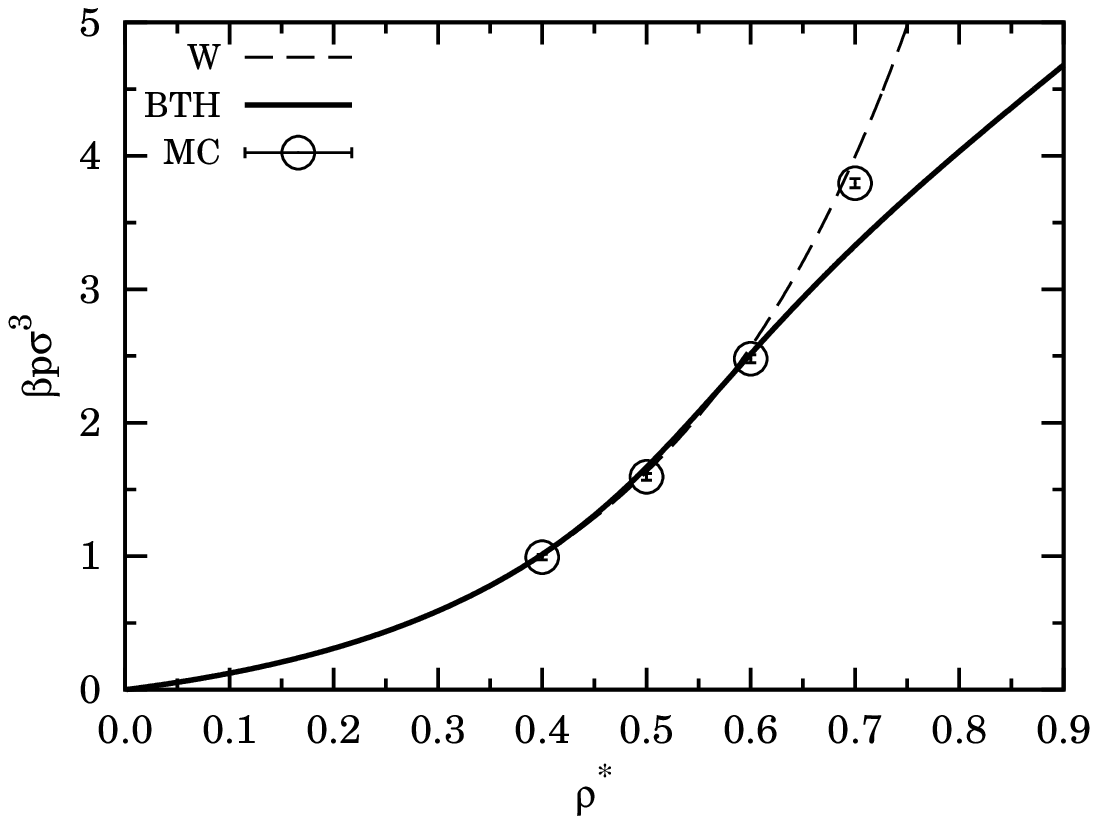}\\
\includegraphics[width=10cm]{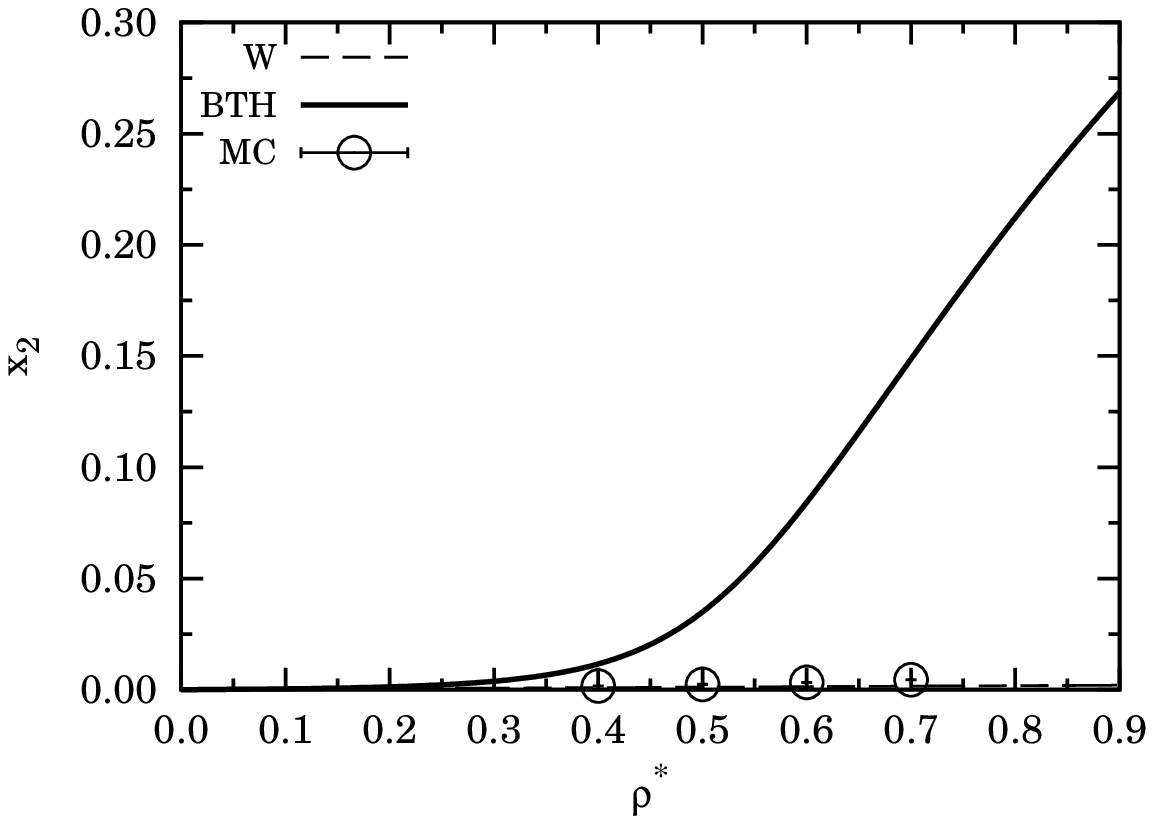}
\end{center}  
\caption{Pressure (top panel) and dimers concentration (bottom panel)
  as a function of density on the $T^*=0.4$ isotherm for ${\cal
    W}=0.1\sigma$ and $\theta_0=\pi/12$. The broken line is the
  prediction 
  of W theory, the continuous line the one of BTH theory with
  $\sigma_c=\sigma$, and the points are the exact MC data.}
\label{fig:pr_wt-kf-t0.4}
\end{figure}

In Figs. \ref{fig:pr_wt-kf-t0.1} and \ref{fig:pr_wt-kf-t0.4} we
compare the simulation data on two different isotherms, at low
temperature $T^*=0.1$ and high temperature $T^*=0.4$, with the dimers
concentrations, $x_2^W$ and $x_2^{BTH}$, and pressures, $\rho z^W$ and
$\rho z^{BTH}$, predicted by Wertheim and BTH theories as shown in
Section \ref{sec:M=1}. From the comparison emerges that at low
temperatures one can adjust $\sigma_c$ in the BTH theory to obtain
good agreement either with the pressure or with the dimers
concentration data, but not with both simultaneously. In the high
temperature limit the two theories coincide for $\sigma_c=\sigma$, but
again BTH fails at high densities at large but finite temperature.


For this system we also tried to use in the BTH theory an intercluster
partition function derived from the Freasier et al. \cite{Binder79}
equation of state for dumbbells with a center-to-center distance equal
to $\sigma$. But we soon discovered that such an equation of state is
very similar to a Carnahan-Starling with a $\sigma_c\approx
2.5\sigma$. This implied that we could study only a density range
$\rho^*<6\sigma^3/(\pi\sigma_c^3)\approx 0.1222$. At such low
densities the fluid tends to dissociate into monomers and as a
consequence such refined BTH becomes worst than the usual BTH with a
Carnahan-Starling intercluster partition function with $\sigma_c$
close to $\sigma$. 
   
%%%%%%%%%%%%%%%%%%%%%%%%%%%%%%%%%%%%%%%%%%%%%%%%%%%%%%%%%%%%%%%%%%%%%%%%%%%%%%
\subsection{Number of cluster species $n_c>2$}
%%%%%%%%%%%%%%%%%%%%%%%%%%%%%%%%%%%%%%%%%%%%%%%%%%%%%%%%%%%%%%%%%%%%%%%%%%%%%%
\label{sec:nc>2}

We have seen in various ways that as long as $n_c\le 2$ we expect,
either from the Wertheim theory or from the BTH theory, the absence of
the liquid phase. So now we want to understand if there exist a
critical $n_c$, $\bar{n}_c$, such that for $n_c>\bar{n}_c$ we may have
the appearance of the liquid in the associating fluid.

According to Wertheim \cite{Wertheim1}: ``{\sl As long as [$n_c$] is
finite, or at least a reasonably small number, we would expect
increasing association with decreasing $T$, but no gas-liquid
transition. On this basis one may conjecture that the gas-liquid
transition is related to the catastrophic increase with $s$ of allowed
$s$-mer[s] [...] when no cutoff [...] is provided.}''.  

Wertheim also suggests that, releasing the single bond per site
condition, a pair-potential of the form given by
Eqs. (\ref{wertheim-potential})-(\ref{site-site-square-well}) allows
to have fluids with $n_c>2$ finite. If Wertheim is correct we would be
unable to predict the liquid phase within the BTH theory. 

In order to understand better this point we looked if it is possible
to have the appearance of a van der Waals loop in
$\beta p^{BTH}=\rho z^{BTH}=\rho^2\partial\beta a^{BTH}/\partial\rho$
for $n_c>2$. 
We looked then at the low temperature $T\to 0$ and large number
of cluster species $n_c\to\infty$ limit. We choose the $z_i\to\infty$
for $i>1$ in the low temperature limit, in such a way to fulfill
complete association, {\sl i.e.} $\lim_{T\to 0}x_{n_c}=1/n_c$. Specifically
we realized this by the choice $z_i=(z_2)^{i-1}$, which can be
justified from the extensive property of the intra-cluster excess
free energy. Then, due to the complete association, we have 
\bq
x_c\stackrel{T\to 0}{\longrightarrow}\frac{1}{n_c}
\stackrel{n_c\to\infty}{\longrightarrow}0,
\eq
so $a_c^{ex}\to 0$. Moreover, it is easy to see, either from a
numerical analysis or analytically, that  
\bq
-\rho<\alpha(n_c)=\lim_{T\to 0}\rho^2
\frac{\partial\left[x_c\ln x_1+(1-x_c)\ln(\lambda e/\rho)\right]}
{\partial\rho}\le -\frac{\rho}{2},
\eq
with $\alpha(n_c)=(1/n_c-1)\rho$ (remember that $\lim_{T\to
  0}\lambda=0$ and temperature and density are two independent
variables) and $\lim_{n_c\to\infty}\alpha(n_c)=-\rho$ and
$\alpha(2)=-\rho/2$ (see Appendix \ref{app:1}). So that, in particular,  
\bq 
\lim_{n_c\to\infty}\lim_{T\to 0}p^{BTH}=0.
\eq
This result strongly suggests that BTH is never able to account for
the liquid phase, contrary to the Wertheim theory
\cite{Bianchi2006,Russo2011a,Rovigatti2013}.  

%%%%%%%%%%%%%%%%%%%%%%%%%%%%%%%%%%%%%%%%%%%%%%%%%%%%%%%%%%%%%%%%%%%%%%%%%%%%%%
\section{Conclusions}
%%%%%%%%%%%%%%%%%%%%%%%%%%%%%%%%%%%%%%%%%%%%%%%%%%%%%%%%%%%%%%%%%%%%%%%%%%%%%%
\label{sec:conclusions}

We compared Wertheim and BTH association theories. Whereas Wertheim
theory is able to account for fluids with an infinite number of
cluster species, BTH is not. As a result, only Wertheim's approach 
is able to account for the percolation and the condensation phenomena. 

For the special case of fluids allowing for dimerization only,
Wertheim theory becomes equivalent to BTH provided an ideal gas
description of the inter-cluster partition function is used. For the
Bjerrum-Tani-Henderson theory we also rigorously proved  
the uniqueness of the solution for the cluster's concentrations and
the reduction of the system of equations to a single one for a single
unknown.

To assess the accuracy of Wertheim and the full BTH using a
hard-sphere (Carnahan-Starling) description of the inter-cluster
partition function, we performed some MC simulations of two dimerizing
systems: a binary mixture of associating non-additive hard-spheres and
a one component single patch Kern-Frenkel fluid. Our results show
that the parameter free Wertheim's theory captures well, at low density, 
the behavior
of the MC data, both for the pressure and the  concentration of dimers,
and the range of densities where it is valid increases with increasing
temperature. BTH, on the other hand, has the dimer diameter as a free
parameter which can be adjusted to find more accurate agreement with the
simulation data, even if the breakdown of its validity at high density
still remains.

\appendix
%%%%%%%%%%%%%%%%%%%%%%%%%%%%%%%%%%%%%%%%%%%%%%%%%%%%%%%%%%%%%%%%%%%%%%%%%%%%%%
\section{Low temperature limit of BTH and W theories} 
%%%%%%%%%%%%%%%%%%%%%%%%%%%%%%%%%%%%%%%%%%%%%%%%%%%%%%%%%%%%%%%%%%%%%%%%%%%%%%
\label{app:1}

For the case studied in Section \ref{sec:results-w-h-1}, from W theory
we find, for the compressibility factor,
\bq
z_{bond}^{W}=\rho\frac{\partial\beta
  a_{bond}^{W}}{\partial\rho}=
-\frac{\Delta\rho}{\left(1+\sqrt{1+2\Delta\rho}\right)^2},
\eq
so, in the low temperature limit, we have
\bq
\lim_{\Delta\to\infty}z_{bond}^{W}=-1/2.
\eq

In BTH theory instead 
\bq
z_{bond}^{BTH}=\rho\frac{\partial\beta
  a_{bond}^{BTH}}{\partial\rho},
\eq
Recalling that $x_c=(1+\lambda z_2)/(1+2\lambda z_2)$, we
find, in the low temperature limit,
$\lim_{z_2\to\infty}x_c=1/2$. Then, for $\sigma_c=\sigma$, we have
$a_c^{ex}\to a_0^{ex}$. So, since $z_2$ and $\rho$ are independent
variables, we find 
\bq
\lim_{z_2\to\infty}z_{bond}^{BTH}=\lim_{z_2\to\infty}
\rho\frac{\partial\left[x_c\ln x_1+(1-x_c)\ln(\lambda
    e/\rho)\right]}{\partial\rho}.
\eq
Observing further that $\lim_{z_1\to\infty}\lambda=0$ we then find
$\lim_{z_2\to\infty}z_{bond}^{BTH}=-1/2$ as for Wertheim.

%%%%%%%%%%%%%%%%%%%%%%%%%%%%%%%%%%%%%%%%%%%%%%%%%%%%%%%%%%%%%%%%%%%%%%%%%%%%%% 
\begin{acknowledgments}
R.F. would like to acknowledge the use of the PLX computational
facility of CINECA through the ISCRA grant. G.P. acknowledges  
financial support by PRIN-COFIN 2010-2011 (contract
2010LKE4CC). It is also a pleasure for G.P. to acknowledge help by
Ms Serena Alfarano to check the proof that it is always possible to
reduce BTH equations to a single equation. 
\end{acknowledgments}
%%%%%%%%%%%%%%%%%%%%%%%%%%%%%%%%%%%%%%%%%%%%%%%%%%%%%%%%%%%%%%%%%%%%%%%%%%%%%%
\bibliographystyle{apsrev}
%\bibliography{association}

\begin{thebibliography}{40}
\bibitem[{not({\natexlab{a}})}]{note0}
\bibinfo{note}{In fact $dx_c/d\lambda<0$ if
  $\sum_{ij}ij\lambda^{i+j-1}z_iz_j<\sum_{ij}i^2\lambda^{i+j-1}z_iz_j$. This
  can be proven observing that for $i\neq j$ we have $(i-j)^2>0$ or $2ij<ii+jj$
  or $ija_{ij}+jia_{ji}<iia_{ij}+jja_{ji}$ for any $a_{ij}=a_{ji}$. Then we
  must also have $\sum_{ij}ija_{ij}<\sum_{ij}iia_{ij}$.}
\bibitem[{not({\natexlab{b}})}]{note1}
\bibinfo{note}{Note that the reference system, the one without attractions and
  therefore with only monomers, will in general differ in the two theories if
  $\sigma_c\neq\sigma$.}
\end{thebibliography}

%%%%%%%%%%%%%%%%%%%%%%%%%%%%%%%%%%%%%%%%%%%%%%%%%%%%%%%%%%%%%%%%%%%%%%%%%%%%%%
%%%%%%%%%%%%%%%%%%%%%%%%%%%%%%%%%%%%%%%%%%%%%%%%%%%%%%%%%%%%%%%%%%%%%%%%%%%%%%
%%%%%%%%%%%%%%%%%%%%%%%%%%%%%%%%%%%%%%%%%%%%%%%%%%%%%%%%%%%%%%%%%%%%%%%%%%%%%%
\end{document}